\documentclass[conference,letterpaper]{IEEEtran}
\IEEEoverridecommandlockouts

\usepackage{cite}
\usepackage{amsmath,amssymb,amsfonts}
\usepackage{algorithmic}
\usepackage{graphicx}
\usepackage{textcomp}
\usepackage{xcolor}
\usepackage{pdfpages}

\def\BibTeX{{\rm B\kern-.05em{\sc i\kern-.025em b}\kern-.08em
    T\kern-.1667em\lower.7ex\hbox{E}\kern-.125emX}}
\usepackage{amsthm}

\theoremstyle{definition}
\newtheorem{thm}{Theorem}

\theoremstyle{definition}

\theoremstyle{lemma}
\newtheorem{lemma}{Lemma}

\theoremstyle{proposition}
\newtheorem{proposition}{Proposition}

\theoremstyle{remark}
\newtheorem{remark}{Remark}
\begin{document}

\title{On Geometric Asymmetry and Information in Sequential Dimension Reduction}

\author{\IEEEauthorblockN{Nazanin Mirhosseini}
\IEEEauthorblockA{\textit{Department of Electrical and Computer Engineering} \\
\textit{Colorado State University}, \\ Fort Collins, CO, USA \\
nazmir@colostate.edu}}

\maketitle

\begin{abstract}
	Standard random projection techniques operate as a black box, mapping high-dimensional structures directly to a lower-dimensional space where the target dimension must be specified a \textit{priori}. To address scenarios where the ultimate dimension is unknown, this paper investigates the retention of information through a sequential dimension reduction process. We examine a fixed, bounded convex body as it undergoes successive random orthogonal projections, systematically reducing the ambient dimension by one at each step. By demonstrating that this sequence of bodies forms a Markov chain, we quantify the information preserved through these reductions using the conditional mutual information between successive projections given the original convex body. We derive an upper bound on this conditional mutual information, parameterized by the Haar measure of the projection spaces that yield the same observed body. Leveraging the established Markov property, we extend these results to an arbitrary number of iterations, proving that the initial two-step bound characterizes information retention across the entire sequence of projections. Furthermore, analyzing the initial body's symmetry group projections reveals that geometric asymmetry enhances information retention.
\end{abstract}



\section{Introduction}\label{sec:intro}

Dimensionality reduction (DR) maps high-dimensional data into a lower-dimensional space while preserving task-relevant information \cite{Jollif2002}, though what is preserved varies: variance in leading eigenspaces for PCA \cite{Jollif2002}, local geometric and topological structure for UMAP \cite{UMAP2020}, class separability for LDA \cite{Fisher1936, ElemStat}, and mutual information with task-relevant variables for information-theoretic methods \cite{Ozdenizci2021, Namekawa2022}.

Random projections provide a data-agnostic alternative with rigorous guarantees \cite{JL2001}. The Johnson-Lindenstrauss (JL) lemma shows that pairwise geometry can be approximately preserved with high probability \cite{JL1988}, and sparse and fast variants reduce storage and computational cost through structured randomness \cite{Kane2014, Ailon2009}. In this paper, we consider an iterative random projection scheme generated from unit vectors on the hypersphere. The resulting mapping is related to QR-type projections \cite{Mezzadri2007}, but our focus is on the stochastic evolution of information across successive projection steps rather than the linear algebraic implementation. This sequential view turns compression into a monitored process rather than a one-shot embedding. At each stage, the remaining information can be quantified to determine whether further projection is warranted. This is especially useful when the target dimension is not known in advance, since it allows adaptive stopping once a prescribed information threshold is reached.

From an information-theoretic standpoint, each projection acts as a channel that compresses a source into a lower-dimensional output \cite{Cover2006}. The central question is thus not only how much geometric distortion is introduced, but how much information survives successive projections. This viewpoint is relevant in resource-constrained inference, where high-dimensional data must be compressed before transmission, and in privacy-preserving representation learning, where randomization is used to obscure identifying structure while retaining utility. Iterative projection schemes provide a principled way to quantify this tradeoff and to relate stopping behavior to information preservation.

For geometric and probabilistic reasons, we model the dataset as a convex body $K_0 \subset \mathbb{R}^n$. Convexity is preserved under linear projection, and the convex hull provides a stable envelope for the data. High-dimensional probability mass often concentrates near a thin shell, and convex measures exhibit sharp information concentration \cite{Vershynin2018, Li2016}. This framework makes it possible to quantify how the symmetry of the original body affects information retention under iteration. Symmetric bodies tend to lose identifying structure more rapidly under random projections, whereas asymmetric bodies retain more distinguishing features across intermediate stages.


\section{Problem Formulation} \label{Sec1}

\subsection{Generating Random Subspaces}

To generate a sequence of random subspaces whose dimension decreases by one at each step, we proceed iteratively. At the first step, let $\boldsymbol{U}_1$ be chosen uniformly at random from the surface of the unit sphere $\mathbb{S}^{n-1}$ in $\mathbb{R}^n$. We then define the orthogonal complement corresponding to $\mbox{\boldmath $U$}_{1}$ by
\begin{equation}
	\mathcal{U}_1^\perp \triangleq \left\{ \boldsymbol{x} \in \mathbb{R}^n : \langle \boldsymbol{x}, \boldsymbol{U}_1 \rangle = 0 \right\}.
\end{equation}
The subspace $\mathcal{U}_1^\perp$ is a random $(n-1)$-dimensional subspace of $\mathbb{R}^n$ and is therefore isomorphic to $\mathbb{R}^{n-1}$ (or $\mathcal{U}_{1}^{\perp} \cong \mathbb{R}^{n-1}$). We denote a vector space isomorphism by $\cong$.  

At the second step, we choose $\boldsymbol{U}_2$ uniformly at random from the unit sphere in $\mathcal{U}_1^\perp$ (equivalently, from $\mathbb{S}^{n-2}$ after identifying $\mathcal{U}_1^\perp \cong \mathbb{R}^{n-1}$), and define its orthogonal complement as
\begin{equation}
\mathcal{U}_2^\perp = \left\{ \boldsymbol{x} \in \mathcal{U}_1^\perp : \langle \boldsymbol{x}, \boldsymbol{U}_2 \rangle = 0 \right\}.
\end{equation}
Then $\mathcal{U}_2^\perp$ is a random subspace of dimension $n-2$.

Continuing in this manner for a $1\leq m < n$, after the $m$-th step  we obtain
\begin{equation}
\mathcal{U}_m^\perp = \left\{ \boldsymbol{x} \in \mathcal{U}_{m-1}^\perp \cong \mathbb{R}^{n-m+1} : \langle \boldsymbol{x}, \boldsymbol{U}_m \rangle = 0 \right\},
\end{equation}
where $\boldsymbol{U}_m$ is uniformly distributed on the unit sphere $\mathbb{S}^{n-m}$ in $\mathcal{U}_{m-1}^\perp$. Hence, $\mathcal{U}_m^\perp$ is a random subspace of dimension $n-m$.  

\subsection{Iterative Projection of a Convex Body onto Random Subspaces}

Let ${K_0} \subset \mathbb{R}^{n}$ be a given bounded convex body\footnote{Throughout this paper, a convex body in Euclidean space $\mathbb{R}^{d}$ is a compact convex set with non-empty interior \cite{Schneider2013}. }. We denote the orthogonal projection of $K_0$ onto $\mathcal{U}_{1}^{\perp}$ by
\begin{equation}
    K_1 \triangleq \mathrm{Proj}_{\mathcal{U}_{1}^{\perp}} (K_0) = \left\{\boldsymbol{x} - \langle \boldsymbol{x}, \boldsymbol{U}_{1}\rangle \boldsymbol{U}_{1} : \boldsymbol{x} \in K_0\right\}.
\end{equation}
Here, $K_1$ should be understood as a convex body lying in the random subspace $\mathcal{W}_1 \triangleq \mathcal{U}_{1}^{\perp} \subset \mathbb{R}^{n}$, which has dimension $n-1$. We next project $K_1$ onto $\mathcal{U}_2^{\perp}$, denoted by
\begin{equation}
    \begin{split}
        K_2 &\triangleq \mathrm{Proj}_{\mathcal{U}_2^{\perp}}(K_1) = \left\{\boldsymbol{x} - \langle \boldsymbol{x}, \boldsymbol{U}_2 \rangle \boldsymbol{U}_2 : \boldsymbol{x} \in K_1\right\} \\
        &= \mathrm{Proj}_{\mathcal{W}_2}(K_0),
    \end{split}
\end{equation}
where $\mathcal{W}_2 \triangleq \mathcal{U}_{1}^{\perp} \cap \mathcal{U}_{2}^{\perp}$. Continuing these projections, the convex body at the $m$-th iteration is
\begin{equation}
    K_m \triangleq \mathrm{Proj}_{\mathcal{U}_m^{\perp}} (K_{m-1}) = \mathrm{Proj}_{\mathcal{W}_m}(K_0),
\end{equation}
where $\mathcal{W}_m \triangleq \bigcap_{i=1}^{m} \mathcal{U}_{i}^{\perp} \subset \mathbb{R}^{n}$. Consequently, $K_m$ is a convex body residing in the $(n-m)$-dimensional random subspace $\mathcal{W}_m$. Note that when we write $K_m \subset \mathbb{R}^{n-m}$, this is only after fixing an orthogonal identification $\mathcal{W}_m \cong \mathbb{R}^{n-m}$.

To ensure the well-definedness of the subsequent information-theoretic analysis, specifically regarding the mutual information $I(K_1; K_2 | K_0)$ and related quantities, we must establish the probabilistic validity of the iterative projections. Formally, the sequence of projected bodies $\{K_i\}_{i=1}^{m}$ must be treated as a sequence of random elements. This follows from the fact that the orthogonal projection map, acting on the space of compact convex sets equipped with the Hausdorff metric, is measurable with respect to the Borel $\sigma$-field. Such measurability ensures that each $K_i$ is a well-defined random variable, justifying the use of probabilistic measures in our analysis.

\begin{lemma}(Measurability of Sequential Projections) \label{lemma1}
	Fix a convex body $K_0 \subset \mathbb{R}^{n}$. Let $T_{K_0}: \mathbb{S}^{n-1} \to \mathcal{K}_{n-1}$ be a mapping between the measure spaces $(\mathbb{S}^{n-1}, \mathcal{B}(\mathbb{S}^{n-1}))$ and $(\mathcal{K}_{n-1}, \mathcal{B}(\mathcal{K}_{n-1}))$, where $\mathbb{S}^{n-1} = \{ \boldsymbol{x} \in \mathbb{R}^n : \|\boldsymbol{x}\|_2 = 1 \}$ is the unit sphere and $\mathcal{K}_{n-1}$ is the set of all convex bodies in $\mathbb{R}^{n-1}$. Here, $\mathcal{B}(\mathbb{S}^{n-1})$ and $\mathcal{B}(\mathcal{K}_{n-1})$ denote the respective Borel $\sigma$-fields, with the latter induced by the Hausdorff metric $d_H$. The map defined by $T_{K_0}(\boldsymbol{u}) = \{ \boldsymbol{x} - \langle \boldsymbol{x}, \boldsymbol{u} \rangle \boldsymbol{u} : \boldsymbol{x} \in K_0 \}$
for any $\boldsymbol{u} \in \mathbb{S}^{n-1}$ is measurable.
\end{lemma}
\begin{proof}
	To prove that the mapping $T_{K_0}$ is measurable for a fixed convex body $K_0$, we prove that this mapping is continuous with respect to Hausdorff metric. To this end, let $\mbox{\boldmath $u$}, \mbox{\boldmath $v$} \in \mathbb{S}^{n-1}$, and fix a point $\mbox{\boldmath $x$}$ from $K_0$, then it yields
	\begin{eqnarray}
		 && \!\!\!\!\!\!\!\! \left \Vert T_{\mbox{\scriptsize \boldmath $x$}\in K_0}(\mbox{\boldmath $u$}) - T_{\mbox{\scriptsize \boldmath $x$}\in K_0}(\mbox{\boldmath $v$}) \right\Vert_2 = \left\| \langle \mbox{\boldmath $x$}, \mbox{\boldmath $v$}\rangle \mbox{\boldmath $v$} - \langle \mbox{\boldmath $x$}, \mbox{\boldmath $u$}\rangle \mbox{\boldmath $u$}\right\|_2\label{substitution}\\
		&& \!\!\!\!\!\!\!\! = \left\| \langle \mbox{\boldmath $x$}, \mbox{\boldmath $u$} \rangle (\mbox{\boldmath $v$}-\mbox{\boldmath $u$}) + \langle \mbox{\boldmath $x$}, \mbox{\boldmath $v$} - \mbox{\boldmath $u$}\rangle  \mbox{\boldmath $v$}\right\|_2 \label{add_sub} \\
		&& \!\!\!\!\!\!\!\! < \left \vert \langle \mbox{\boldmath $x$},\mbox{\boldmath $u$}\rangle\right \vert \| \mbox{\boldmath $v$}-\mbox{\boldmath $u$}\|_2 + | \langle \mbox{\boldmath $x$}, \mbox{\boldmath $v$}-\mbox{\boldmath $u$}\rangle | \| \mbox{\boldmath $v$}\|_2 \label{triangle_ineq}\\
		&& \!\!\!\!\!\!\!\! < 2\| \mbox{\boldmath $x$}\|_2 \| \mbox{\boldmath $v$}-\mbox{\boldmath $u$}\|_2  \label{out_triangleq} \stackrel{(a)} {<} 2R \| \mbox{\boldmath $v$} - \mbox{\boldmath $u$}\|_2 ,
	\end{eqnarray}
	where (\ref{add_sub}) is by adding and subtracting $\langle \mbox{\boldmath $x$}, \mbox{\boldmath $u$}\rangle \mbox{\boldmath $v$}$ to (\ref{substitution}), (\ref{triangle_ineq}) follows from applying triangle inequality to (\ref{add_sub}) and (\ref{out_triangleq}) is by $| \langle \mbox{\boldmath $a$}, \mbox{\boldmath $b$} \rangle | \leq \| \mbox{\boldmath $a$}\|_2 \| \mbox{\boldmath $b$}\|_2$ for $\mbox{\boldmath $a$},\mbox{\boldmath $b$}\in \mathbb{R}^{n}$, and this fact that $\| \mbox{\boldmath $v$}\|_2 = \| \mbox{\boldmath $u$}\|_2 = 1$. Here, since $K_0$ is a compact set, we have $R = \sup_{\mbox{\scriptsize \boldmath $x$}\in K_0} \| \mbox{\boldmath $x$}\|_2 < \infty$, from which $(a)$ is followed. Hence, for any $T_{\mbox{\scriptsize \boldmath $x$}\in K_0}(\mbox{\boldmath $u$}) \in T_{K_0}(\mbox{\boldmath $u$})$, there exists $T_{\mbox{\scriptsize \boldmath $x$}\in K_0}(\mbox{\boldmath $v$}) \in T_{K_0}(\mbox{\boldmath $v$})$ such that $\left \| T_{\mbox{\scriptsize \boldmath $x$}\in K_0}(\mbox{\boldmath $u$}) - T_{\mbox{\scriptsize \boldmath $x$}\in K_0}(\mbox{\boldmath $v$})\right \|_2 < 2R \| \mbox{\boldmath $u$} - \mbox{\boldmath $v$}\|_2,$
	resulting into 
	\begin{eqnarray}
		&& \!\!\!\!\!\!\!\!\!\!\! \sup_{T_{\mbox{\scriptsize \boldmath $x$}\in K_0}(\mbox{\scriptsize \boldmath $u$}) \in T_{K_0}(\mbox{\scriptsize \boldmath $u$})} \inf_{T_{\mbox{\scriptsize \boldmath $x$}\in K_0}(\mbox{\scriptsize \boldmath $v$}) \in T_{K_0}(\mbox{\scriptsize \boldmath $v$})} \| T_{\mbox{\scriptsize \boldmath $x$}\in K_0}(\mbox{\boldmath $u$}) - T_{\mbox{\scriptsize \boldmath $x$}\in K_0}(\mbox{\boldmath  $v$}) \|_2 \nonumber \\
		&& \ \ \ \ \ \ \ \ \ \ \ \ \ \ \ \ \ \ \ \ \ \ \ \ \ \ \ \ \ \ \ \ \ \ \ \ < 2R\| \mbox{\boldmath $u$} - \mbox{\boldmath $v$}\|_2. \label{Hausdorff1}
	\end{eqnarray}
	Similarly, 
	\begin{eqnarray}
		&&  \!\!\!\!\!\!\!\!\!\!\! \sup_{T_{\mbox{\scriptsize \boldmath $x$}\in K_0}(\mbox{\scriptsize \boldmath $v$}) \in T_{K_0}(\mbox{\scriptsize \boldmath $v$})} \inf_{T_{\mbox{\scriptsize \boldmath $x$}\in K_0}(\mbox{\scriptsize \boldmath $u$}) \in T_{K_0}(\mbox{\scriptsize \boldmath $u$})} \| T_{\mbox{\scriptsize \boldmath $x$}\in K_0}(\mbox{\boldmath $u$}) - T_{\mbox{\scriptsize \boldmath $x$}\in K_0}(\mbox{\boldmath  $v$}) \|_2 \nonumber \\
		&& \ \ \ \ \ \ \ \ \ \ \ \ \ \ \ \ \ \ \ \ \ \ \ \ \ \ \ \ \ \ \ \ \ \ \ \ < 2R\| \mbox{\boldmath $u$} - \mbox{\boldmath $v$}\|_2.\label{Hausdorff2}
	\end{eqnarray}
	According to (\ref{Hausdorff1}),  (\ref{Hausdorff2}), and by definition of Hausdorff distance for all $A, B \in \mathcal{K}_{n-1}$ 
	 \begin{eqnarray}
		d_{H}(A,B)  \triangleq \max \left\{ \sup_{\mbox{\scriptsize \boldmath $a$}\in A}\inf_{\mbox{\scriptsize \boldmath $b$}\in B} \|\mbox{\boldmath $a$}-\mbox{\boldmath $b$}\|_2, \sup_{\mbox{\scriptsize \boldmath $b$}\in B}\inf_{\mbox{\scriptsize \boldmath $a$}\in A} \|\mbox{\boldmath $a$}-\mbox{\boldmath $b$}\|_2 \right\}, \nonumber
	\end{eqnarray} 
	we have $d_{H}\left( T_{K_0}(\mbox{\boldmath $u$}), T_{K_0}(\mbox{\boldmath $v$})\right) < 2R \| \mbox{\boldmath $u$} - \mbox{\boldmath $v$}\|_2$,
	which indicates that the mapping $\mbox{\boldmath $u$} \mapsto T_{K_0}(\mbox{\boldmath $u$})$ is Lipschitz continuous and hence uniformly continuous with  respect to Euclidean metric on $\mathbb{S}^{n-1}$ and Hausdorff distance on $\mathcal{K}_{n-1}$. Since continuous functions are measurable, and uniform continuity implies that $T_{K_0}(\mbox{\boldmath $u$})$ is continuous with respect to Hausdorff distance, $T_{K_0}(\mbox{\boldmath $u$})$ is measurable.
\end{proof} 
	
\section{Upper Bound on Mutual Information}
As established in Lemma \ref{lemma1}, the orthogonal projection of a fixed convex body $K_0 \subset \mathbb{R}^{n}$ onto $\mathcal{U}_{1}^{\perp}$ yields a random variable $K_1$, mapping from $\mathbb{S}^{n-1}$ to $\mathcal{K}_{n-1}$. The reasoning in Lemma \ref{lemma1} can be extended to higher-order iterations to prove that the projected convex body over each iteration is a random variable. Consequently, we can extend the definition of Shannon entropy to random convex bodies.

\begin{thm} \label{MainThm}
	Let $K_0 \subset \mathbb{R}^{n}$ be a fixed convex body. As described, $K_1 = \mathrm{Proj}_{{\mathcal{U}_{1}}^{\perp}}(K_0)$ and $K_2 = \mathrm{Proj}_{{\mathcal{U}_{2}}^{\perp}}(K_1)$. Then, 
	\begin{eqnarray}
		I(K_1;K_2|K_0) \leq \log \frac{\pi^{\frac{n}{2}-2}}{\Gamma\left(\frac{n-2}{2}\right)} - \mathbb{E}_{K_2|K_0}\left[\log  {N}(K_0,K_2)\right], \nonumber
	\end{eqnarray}
	where $\Gamma(.)$ is the Gamma function, 
	\begin{eqnarray}
		&& N(K_0, K_2) = \frac{\mathrm{Vol}(\mathcal{C}(K_0,K_2))}{\mathrm{Vol}(G_{n,2})},\nonumber\\
		&& \mathcal{C}(K_0,K_2) = \{ \mathcal{W}\in G_{n,2}:\mathrm{Proj}_{\mathcal{W}^{\perp}}(K_0) = K_2\},\nonumber
	\end{eqnarray}
	and $\mathrm{Vol}$ is with respect to Haar measure on the Grassmannian manifold of $2$-planes in $\mathbb{R}^{n}$ denoted by $G_{n,2}$. 
\end{thm}

\begin{proof}
	Since $K_1$ is a deterministic function of $\mbox{\boldmath $U$}_{1}$,  by data processing inequality, we have
	\begin{eqnarray}
		&& I(K_1;K_2|K_0) \leq I(K_2;\mbox{\boldmath $U$}_{1}|K_0)  \nonumber \\
		&& \ \ \ \ \ \ \ \ \ \ \ \ \ \ \ \ \ \  = h(\mbox{\boldmath $U$}_{1}|K_0) - h(\mbox{\boldmath $U$}_{1}| K_2,K_0), \label{DefMutual}
	\end{eqnarray}
	where $h(.|.)$ is the conditional differential entropy. Leveraging the fact that $\mbox{\boldmath $U$}_{1} \sim \mathrm{Uniform}\left(\mathbb{S}^{n-1}\right)$, we simplified (\ref{DefMutual}) as the following
	\begin{eqnarray}
		I(K_1;K_2|K_0) = \log \frac{2\pi^{n/2}}{\Gamma\left(\frac{n}{2}\right)}-h(\mbox{\boldmath $U$}_{1}|K_2,K_0). \label{DefMutual2}
	\end{eqnarray}
	Now, we wish to derive a lower bound for $h(\mbox{\boldmath $U$}_{1}|K_2,K_0)$. To this end, denote $\mathcal{W}_2 = {\mathcal{U}_1}^{\perp} \cap {\mathcal{U}_2}^{\perp}$, whose orthogonal complement is ${\mathcal{W}_2}^{\perp} = \mathrm{Span}\{ \mathcal{U}_{1}\cup \mathcal{U}_{2}\}$. Note that $\mathcal{W}_{2}$ is a subspace with codimension $2$. Given $K_2$, we establish the set of codimension-2 subspaces yielding $K_2$, denoted by $\mathcal{C}(K_0,K_2)$ as the following
	\begin{eqnarray}
		\mathcal{C}(K_0,K_2) = \left\{ \mathcal{W} \in  G_{n,2}: \mathrm{Proj}_{\mathcal{W}^{\perp}}(K_0) = K_2 \right\}.
	\end{eqnarray}
	Given $K_0$ and the realization $K_2 = k_2$ for $k_2 \in \mathcal{K}_{n-2}$, the conditional distribution of $\mathcal{W}_{2}$ (a random subspace which is determined by the pair $(\mbox{\boldmath $U$}_{1},\mbox{\boldmath $U$}_{2})$) is uniform over $\mathcal{C}(K_0,k_2)$ with respect to the invariant measure on $G_{n,2}$. The reason is that the process of choosing $(\mbox{\boldmath $U$}_{1}, \mbox{\boldmath $U$}_{2})$ uniformly on the Stiefel manifold induces a uniform distribution on $G_{n,2}$ for $\mathcal{W}_{2}$, and conditioning on $K_2=k_2$ restricts $\mathcal{W}_{2}$ to $\mathcal{C}(K_0,k_2)$. For a fixed $\mathcal{W}_{2}$, since $\mathcal{W}_{2} = {\mathcal{U}_1}^{\perp} \cap {\mathcal{U}_2}^{\perp}$, and conditioning on $K_2=k_2$ results in $\mathcal{W}_{2} \in \mathcal{C}(K_0,k_2)$, the direction $\mbox{\boldmath $U$}_{1}$ must lie on the unit circle in $2$-dimensional subspace ${\mathcal{W}_2}^{\perp}$. Duo to the rotational symmetry, and uniformity of $\mbox{\boldmath $U$}_{1}$ on the sphere, the conditional distribution of $\mbox{\boldmath $U$}_{1}$ given $\mathcal{W}_2$ and $K_2 = k_2$ is uniform on circle. Hence,
	\begin{eqnarray}
		h(\mbox{\boldmath $U$}_{1} | \mathcal{W}_2, K_2=k_2, K_0) = \log 2\pi. \label{h_U_1}
	\end{eqnarray}
	\begin{remark}(\textit{Justifying Conditioning on Measure zero event $\{K_2=k_2\}$}) 
Endowed with the Hausdorff metric, $\mathcal{K}_{n-2}$ is a closed subset of a Polish space of non-empty compact sets. Consequently, both $\mathcal{K}_{n-2}$ and $\mathbb{S}^{n-1}$ are standard Borel spaces, and \cite[Theorem 10.2.2]{Dudley2002} guarantees the existence of a regular conditional distribution $\nu(k_2, \cdot)$. Thus, for any $B \in \mathcal{B}(\mathbb{S}^{n-1})$, we can define $\mathbb{P}[\mbox{\boldmath $U$}_{1} \in B \mid K_2=k_2] = \nu(k_2,B)$ for $P_{K_2}$-almost every $k_2$. Because $\nu(k_2, \cdot)$ is constructed directly from the joint law rather than by dividing by $\mathbb{P}[K_2=k_2]$, the conditional probability remains well-defined even though $\{K_2=k_2\}$ is a measure zero event.
\end{remark}

	By definition of conditional mutual information, we have
	\begin{eqnarray}
		&& \!\!\!\!\!\!\!\!\!\! I(\mbox{\boldmath $U$}_{1};\mathcal{W}_2|K_2,K_0) \!=\! h(\mbox{\boldmath  $U$}_{1}|K_2,K_0)\! -\! h(\mbox{\boldmath $U$}_{1}|\mathcal{W}_2,K_2,K_0). \nonumber
	\end{eqnarray}
	\vspace{-2mm}
	Hence,
	\begin{eqnarray}
		h(\mbox{\boldmath $U$}_{1}|K_2,K_0) \!\!\!\!\!\!\!\!\!\!\! && = h(\mbox{\boldmath $U$}_{1}|\mathcal{W}_2,K_2,K_0) + I(\mbox{\boldmath $U$}_{1};\mathcal{W}_2|K_2,K_0) \nonumber \\ 
		&& = \log 2\pi + I(\mbox{\boldmath $U$}_{1};\mathcal{W}_2|K_2,K_0), \label{substitute_What_we_said_above}
	\end{eqnarray}
	where (\ref{substitute_What_we_said_above}) follows from (\ref{h_U_1}). By definition of conditional mutual information, we expand the mutual information term in (\ref{substitute_What_we_said_above}) as the following
	\begin{eqnarray}
		&& \!\!\!\!\!\!\!\!\!\!\!\!\!\!\!\!\!\!\!\! I\!(\mbox{\boldmath $U$}_{1};\mathcal{W}_2|K_2,K_0\!) \! =\! h(\mathcal{W}_2|K_2,K_0\! )\! - \! h(\mathcal{W}_2|\mbox{\boldmath $U$}_{1}\! ,K_2 ,K_0\! ). \label{mutual2}
	\end{eqnarray}
	Since the conditional distribution of $\mathcal{W}_2$ given $K_2= k_2$, as mentioned earlier, is uniform on $\mathcal{C}(K_0,k_2)$, then $h(\mathcal{W}_2| K_2=k_2,K_0) = \log \mathrm{Vol}(\mathcal{C}(K_0,k_2))$, from which 
	\begin{eqnarray}
		h(\mathcal{W}_2|K_2,K_0) = \mathbb{E}_{K_2|K_0} \left[\log \mathrm{Vol}(\mathcal{C}(K_0,K_2))\right]. \label{firstpart}
	\end{eqnarray}
	To upper bound $h(\mathcal{W}_2|\mbox{\boldmath $U$}_{1},K_2,K_0)$, we proceed as the following. Given $\mbox{\boldmath $U$}_{1}$ and $K_2=k_2$, according to $\mathcal{W}_2 = {\mathcal{U}_1}^{\perp} \cap {\mathcal{U}_2}^{\perp}$, we know that $\mathcal{W}_2 \subseteq {\mathcal{U}_1}^{\perp}$ and $\mathcal{W}_2 \in \mathcal{C}(K_0,k_2)$. The set of possible $\mathcal{W}_2$ given $\mbox{\boldmath $U$}_{1}$ is contained in Grassmannian manifold $G_{n-1,2}$ (codimension-2 subspaces within the hyperplane  ${\mathcal{U}_1}^{\perp} \cong \mathbb{R}^{n-1}$). Therefore,
	\begin{eqnarray}
		h(\mathcal{W}_2|\mbox{\boldmath $U$}_{1},K_2,K_0) \leq \log \mathrm{Vol}(G_{n-1,2}).\label{secondpart}
	\end{eqnarray}
	According to (\ref{mutual2}), by combining (\ref{substitute_What_we_said_above}), (\ref{firstpart}), and (\ref{secondpart}), we have
	\begin{eqnarray}
		&& \!\!\!\!\!\!\!\!\!\!\!\!\!\! h(\mbox{\boldmath $U$}_{1}|K_2,K_0) \geq \log 2\pi + \mathbb{E}_{K_2|K_0}\left[\log \mathrm{Vol}(\mathcal{C}(K_0,K_2)\right]\label{Boundonh}\\
		&& \ \ \ \ \ \ \ \ \ \ \ \ \ \ \ \ \ \ \ \ \ \ \ \ \ \ \ \ \ \ \ \ \ \ \ \ \ - \log \mathrm{Vol}(G_{n-1,2}).\nonumber
	\end{eqnarray}
	To measure the relative volume of subspaces $\mathcal{W}_2$ that produce the same projection $K_2$ when $K_0$ is projected onto them, we define $N(K_0,K_2)$ as the following
	\begin{eqnarray}
		N(K_0,K_2) \triangleq \frac{\mathrm{Vol}(\mathcal{C}(K_0,K_2))}{\mathrm{Vol}(G_{n,2})}.\label{N_K_K2}
	\end{eqnarray}
	Using this definition, we rewrite the bound in (\ref{Boundonh}) as follows
	\begin{eqnarray}
		&& h(\mbox{\boldmath $U$}_{1}|K_2,K_0) \geq\nonumber \\
		&& \log 2\pi + \mathbb{E}_{K_2|K_0}\left[\log N(K_0,K_2)\right] + \log \frac{\mathrm{Vol}(G_{n,2})}{\mathrm{Vol}(G_{n-1,2})}\nonumber \\
		&& = \log 2\pi^2 \frac{\Gamma\left(\frac{n-2}{2}\right)}{\Gamma\left(\frac{n}{2}\right)} + \mathbb{E}_{K_2|K_0}\left[\log N(K_0,K_2)\right], \label{VolG}
\end{eqnarray}	 
where (\ref{VolG}) is by substituting the volume 
\vspace{-3pt}
\begin{eqnarray}
	\mathrm{Vol}(G_{n,2}) = \frac{4\pi^{n-\frac{1}{2}}}{\Gamma\left(\frac{n}{2}\right)\Gamma\left(\frac{n-1}{2}\right)}.
\end{eqnarray} 

By combining (\ref{DefMutual2}) and (\ref{VolG}), we have 
	\begin{eqnarray}
		&& \!\!\!\!\!\!\! I(K_1;K_2|K_0) \leq \nonumber \\
		&& \!\!\!\!\!\!\! \log \frac{2\pi^{\frac{n}{2}}}{\Gamma\left(\frac{n}{2}\right)} - \log 2\pi^2 \frac{\Gamma\left(\frac{n-2}{2}\right)}{\Gamma\left(\frac{n}{2}\right)} - \mathbb{E}_{K_2|K_0}\left[\log  {N}(K_0,K_2)\right]\nonumber\\
		&& \!\!\!\!\!\!\! =\log \frac{\pi^{\frac{n}{2}-2}}{\Gamma\left(\frac{n-2}{2}\right)} - \mathbb{E}_{K_2|K_0}\left[\log  {N}(K_0,K_2)\right],
	\end{eqnarray}
	which completes the proof.
\end{proof}

\begin{remark}
Note that $N(K_0,K_2)$ quantifies the ambiguity in subspace $\mathcal{W}_2$ given a specific projection $K_2$. By definition of $N(K_0,K_2)$ in (\ref{N_K_K2}), we know that $0<N(K_0,K_2)<1$ almost surely. As $N(K_0,K_2)$ becomes closer to $0$, it represents low ambiguity i.e., fewer subspaces $\mathcal{W}_2$ yield $K_2$, and as $N(K_0,K_2)$ becomes closer to $1$, many subspaces $\mathcal{W}_2$ yield $K_2$ which represents higher ambiguity.
\end{remark}

\begin{remark}
The normalized Haar measure on $G_{n,2}$ by $\mathrm{Vol}(G_{n,2})$ is an invariant probability measure. The pushforward of this Haar measure under the map $\mathcal{W} \mapsto \mathrm{Proj}_{\mathcal{W}^{\perp}}(K_0)$ induces a probability measure on $\mathcal{K}_{n-2}$ (space of convex bodies in  $\mathbb{R}^{n-2}$). For a fixed initial convex body $K_0$, the probability that the random projection $\mathrm{Proj}_{\mathcal{W}^{\perp}}(K_0)$ equals $K_2$ is the measure of $\mathcal{C}(K_0,K_2)$. Hence,
	\begin{eqnarray}
		\mathbb{P}\left[\mathrm{Proj}_{\mathcal{W}^{\perp}}(K_0) \! = \! K_2 \middle \vert K_0 \right] && \!\!\!\!\!\!\!\!\!\! =\! \frac{\mathrm{Vol}(\mathcal{C}(K_0,K_2))}{\mathrm{Vol}(G_{n,2})}\! = \! N(K_0,K_2)\nonumber ,
	\end{eqnarray}
	which is the probability density with respect to pushforward measure of Haar measure on $G_{n,2}$.
\end{remark}


\begin{lemma} \label{MarkovChain}
Let $K_0 \subset \mathbb{R}^n$ be a given bounded convex body. The sequence of random subspaces is generated iteratively as described. Then the stochastic process $\{K_m\}_{m=1}^{n-1}$ forms a time-inhomogeneous Markov chain on the space of convex bodies.
\end{lemma}

\begin{proof}
For each $m$, the convex body $K_m$ lies in the random subspace $\mathcal{W}_m$. Since orthogonal projection preserves convexity and $K_0$ is full-dimensional, the affine hull of $K_m$ (the smallest affine subspace containing $K_m$) coincides with $\mathcal{W}_m$, namely, $\mathrm{aff}(K_m)=\mathcal{W}_m$. Hence, the subspace $\mathcal{W}_m$ is completely determined by the present state $K_m$. At step $m+1$, the random vector $\boldsymbol{U}_{m+1}$ is chosen uniformly on the unit sphere contained in $\mathcal{W}_m$, and $\mathcal{W}_{m+1}= \mathcal{W}_m \cap \mathcal{U}_{m+1}^\perp$.
The next convex body is therefore given by $K_{m+1}=\mathrm{Proj}_{\mathcal{W}_{m+1}}(K_0)$.
Using the nesting relation $\mathcal{W}_{m+1}\subset \mathcal{W}_m$ together with the composition property of orthogonal projections,
\begin{equation}
	\mathrm{Proj}_{\mathcal{W}_{m+1}}
\circ
\mathrm{Proj}_{\mathcal{W}_m}
=
\mathrm{Proj}_{\mathcal{W}_{m+1}},
\end{equation}
we obtain $K_{m+1}=\mathrm{Proj}_{\mathcal{W}_{m+1}}(K_m)$.
Thus, conditioned on $K_m$, the random body $K_{m+1}$ depends only on the newly sampled random direction $\boldsymbol{U}_{m+1}$, whose conditional distribution is uniform over the unit sphere in $\mathcal{W}_m=\mathrm{aff}(K_m)$.

Let $\mathcal{F}_m=\sigma(K_1,\ldots,K_m)$ denote the natural filtration. For every measurable set $\mathcal{A}$ of convex bodies,
\begin{equation}
	\mathbb{P}[K_{m+1}\in \mathcal{A}\mid \mathcal{F}_m]
=
\mathbb{P}[K_{m+1}\in \mathcal{A}\mid K_m].
\end{equation}
Hence, the process $\{K_m\}_{m=1}^{n-1}$ satisfies the Markov property and therefore forms a Markov chain.
\end{proof}

\begin{remark}
The Markov chain is time-inhomogeneous since the dimension of the ambient subspace decreases at each iteration, $\dim(\mathcal{W}_m)=n-m$.
Conditioned on $K_m$, the transition kernel is given by
\begin{equation}
	\mathbb{P}[K_{m+1}\in \mathcal{A}\mid K_m]
=
\int_{\mathbb{S}(\mathcal{W}_m)}
\!\!\! \mathbf{1}_{\mathcal{A}}
\!\left\{
\mathrm{Proj}_{\boldsymbol{u}^\perp}(K_m)
\right\}
\, \mathrm{d}\sigma_{\mathcal{W}_m}(\boldsymbol{u}), \nonumber
\end{equation}
where $\sigma_{\mathcal{W}_m}$ denotes the uniform probability measure on the unit sphere in $\mathcal{W}_m$, and $\mathbb{S}(\mathcal{W}_{m})$ is the unit sphere inside the affine hull of $\mathcal{W}_{m}$.
\end{remark}

\begin{proposition}\label{Prop1}
	For a random sequence of convex bodies $K_1,..., K_m$ given $K_0$, we have $I(K_1;K_m|K_0) < I(K_1;K_2|K_0)$.
\end{proposition}
\begin{proof}
	As a direct consequence of Lemma \ref{MarkovChain}, we have $I(K_1;K_3,\ldots, K_m|K_2,K_0) = 0$,
	combining with chain rule, we obtain
	\vspace{-2mm}
	\begin{eqnarray}
		&& \!\!\!\!\!\!\!\!\!\!\!\!\!\!\!\!\!\!\! I(K_1;K_2,K_3,\ldots, K_m|K_0) = I(K_1;K_2|K_0) \nonumber\\
		&& \!\!\!\!\!\!\!\!\!\!\!\!\!\!\!\!\!\!\! + I(K_1;K_3,\ldots, K_m|K_2,K_0) = I(K_1;K_2|K_0).\label{ChainRule1}
	\end{eqnarray}
	Once again by chain rule, it yields
	\begin{eqnarray}
		&& \!\!\!\!\!\!\!\!\!\!\!\!\!\!\!\!\! I(K_1;K_2,K_3,\ldots, K_m|K_0) = I(K_1;K_m|K_0)\nonumber\\
		&& \!\!\!\!\!\!\!\!\!\!\!\!\!\!\!\!\!\!\! + I(K_1;K_2,\ldots, K_{m-1}|K_m,K_0) > I(K_1;K_m|K_0).\label{ChainRule2}
	\end{eqnarray}
	Comparing (\ref{ChainRule1}) with (\ref{ChainRule2}) completes the proof.
\end{proof}

\section{The Effect of Asymmetry on Mutual Information} \label{Symm}

In Theorem \ref{MainThm}, we showed that the bound on $I(K_1;K_2|K_0)$ comprises two terms: one depending only on the dataset dimension $n$, and one depending on dataset structure via $\mathbb{E}_{K_2|K_0}\left[\log N(K_0,K_2)\right]$. By Proposition \ref{Prop1}, this also bounds $I(K_1;K_m|K_0)$. We now wish to relate $N(K_0,K_2)$ to the geometry of $K_0$, particularly its asymmetry.


\begin{thm} \label{ThmEBound}
	Let $K_0 \subset \mathbb{R}^{n}$ be a fixed convex body with finite symmetry group $G = \{g\in O(n): gK_0 = K_0\}$. If we stratify the Grassmannian $G_{n,2}$ by orbit types under action $G$ into the following disjoint decompositions for conjugacy classes $[H] \in \mathcal{S}$,
	\begin{eqnarray}
		G_{n,2}^{[H]} = \{ \mathcal{W}\in G_{n,2}:H_{\mathcal{W}} \ \text{is conjugate to $H \subseteq G$}\},\nonumber
	\end{eqnarray}
	where $H_{\mathcal{W}}$ is the stabilizer subgroup of $G$ for a $\mathcal{W} \in G_{n,2}$, and $\mathcal{S}$ is the set of conjugacy classes of subgroups $H \subseteq G$ which are stabilizers for some  $\mathcal{W}\in  G_{n,2}$. Let $\mu$ be  the uniform probability measure on $G_{n,2}$, then
	\begin{eqnarray}
		&& \!\!\!\!\!\!\!\!\!\!\!\!\!\!\!\!\!\!\! \mathbb{E}_{K_2|K_0}\left[\log  N(K_0,K_2)\right] \geq \nonumber \\
		&& \log \frac{\mathrm{Vol}(G)}{\mathrm{Vol}(G_{n,2})}  - \sum_{[H]\in \mathcal{S}}\log v_{[H]} \mu\left(G_{n,2}^{[H]}\right), \nonumber
	\end{eqnarray}
	where $v_{[H]}$ is a constant.
\end{thm}

\begin{proof}
	Our goal is to find a lower bound for $\mathbb{E}_{K_2|K_0}\left[\log N(K_0,K_2)\right]$, which is dependent on $K_0$ through its corresponding \textit{symmetry group}. Let the symmetry group of $K_0$ be $G \triangleq \left\{g \in O(n): gK_0 = K_0\right\}$,
where $O(n)$ is the orthogonal group in $\mathbb{R}^{n}$. The symmetry group $G$ is the stabilizer subgroup under orthogonal transformations. As mentioned in the proof of Theorem \ref{MainThm}, the distribution of $\mathcal{W}_2$ on $G_{n,2}$ is uniform. Leveraging this fact that the randomness of $K_2$ originates from the randomness of $\mathcal{W}_2$, we have
\begin{eqnarray}
	&& \!\!\!\!\!\!\!\!\!\!\!\!\!\!\!\!\!\! \mathbb{E}_{K_2|K_0}\left[\log N(K_0,K_2)\right] = \nonumber \\
	&& \int_{G_{n,2}}\log \frac{\mathrm{Vol}(\mathcal{C}(K_0,\mathrm{Proj}_{\mathcal{W}^{\perp}}(K_0)))}{\mathrm{Vol}(G_{n,2})}\mathrm{d}\mu(\mathcal{W}), \label{integral1}
\end{eqnarray}
where $\mu$ is the uniform probability measure on $G_{n,2}$. For a fixed $\mathcal{W}$, the set $\mathcal{C}(K_0,\mathrm{Proj}_{\mathcal{W}^{\perp}}(K_0))$ contains orbit of $\mathcal{W}$ under $G$. Thus, $G.\mathcal{W} = \{ g\mathcal{W}: g\in G\} \subseteq \mathcal{C}(K_0,\mathrm{Proj}_{\mathcal{W}^{\perp}}(K_0))$,
which results in
\begin{eqnarray}
	\mathrm{Vol}(G.\mathcal{W}) \leq \mathrm{Vol}(\mathcal{C}(K_0,\mathrm{Proj}_{\mathcal{W}^{\perp}}(K_0))), \label{U1}
\end{eqnarray}
where $\mathrm{Vol}$ is the Haar measure on $G_{n,2}$. Let the stabilizer subgroup of $G$ for a fixed $\mathcal{W}$ be denoted by
\begin{eqnarray}
	H_{\mathcal{W}} \triangleq  \{ g\in G: g\mathcal{W} = \mathcal{W} \}.\label{stabilizer}
\end{eqnarray}
The orbit $G.\mathcal{W}$ is a homogeneous space isomorphism to quotient $G / H_{\mathcal{W}}$. Hence, 
\begin{eqnarray}
	\mathrm{Vol}(G.\mathcal{W}) = \frac{\mathrm{Vol}(G)}{\mathrm{Vol}(H_{\mathcal{W}})},\label{U2}
\end{eqnarray}
with respect to normalized Haar measure on $G_{n,2}$. By definition of $N(K_0,K_2)$ in (\ref{N_K_K2}) combining with (\ref{U1}) and  (\ref{U2}), we obtain
\begin{eqnarray}
	N(K_0,K_2) = \frac{\mathrm{Vol}(\mathcal{C}(K_0,K_2))}{\mathrm{Vol}(G_{n,2})}  \geq \frac{\mathrm{Vol}(G)}{\mathrm{Vol}(H_{\mathcal{W}})\mathrm{Vol}(G_{n,2})}.\label{U3}
\end{eqnarray}
By (\ref{integral1}) and (\ref{U3}), we have
\begin{eqnarray}
	&& \!\!\!\!\!\!\!\!\!\!\!\!\!\!\!\!\!\!\! \mathbb{E}_{K_2|K_0}\left[\log N(K_0,K_2)\right] \geq \nonumber \\
	&& \log \frac{\mathrm{Vol}(G)}{\mathrm{Vol}(G_{n,2})} -  \int_{G_{n,2}}\log \mathrm{Vol}(H_{\mathcal{W}})\mathrm{d}\mu(\mathcal{W}).\label{integral2}
\end{eqnarray}
To compute the integral over $G_{n,2}$ in (\ref{integral2}), we leverage the stratification of Grassmannian $G_{n,2}$ by orbit types under the action of symmetry group $G \subset O(n)$ of convex bodies. This stratification organizes subspaces $\mathcal{W}\in G_{n,2}$ based on the conjugacy class of their stabilizer $H_{\mathcal{W}}=\{g\in  G:g\mathcal{W}= \mathcal{W} \}$. The integral in (\ref{integral2}) is then evaluated by  summing over these strata, where $\mathrm{Vol}(H_{\mathcal{W}})$  is constant on each stratum.

Let $\mathcal{S}$ be the set of conjugacy classes $[H]$ of closed subgroups $H \subseteq G$ such that they are stabilizers, as in (\ref{stabilizer}), for some $\mathcal{W} \in G_{n,2}$. We define a stratum for each conjugacy class $[H] \in \mathcal{S}$ as the following
\begin{eqnarray}
	G_{n,2}^{[H]} \triangleq \{ \mathcal{W} \in G_{n,2} : H_{\mathcal{W}} \ \text{is conjugate to $H \subseteq G$}\},
\end{eqnarray}
which leads to
\begin{eqnarray}
	G_{n,2} = \bigsqcup_{[H] \in \mathcal{S}} G_{n,2}^{[H]}. \label{parition}
\end{eqnarray}
By (\ref{parition}), we rewrite the integral in  (\ref{integral2}) as the following
\begin{eqnarray}
	\int_{G_{n,2}} \!\!\!\!\!\!\! \log \mathrm{Vol}(H_{\mathcal{W}}) \mathrm{d}\mu(\mathcal{W}) \! = \!\!\!  \sum_{[H]\in \mathcal{S}}\int_{G_{n,2}^{[H]}} \!\!\!\! \log \mathrm{Vol}(H_{\mathcal{W}}) \mathrm{d}\mu(\mathcal{W}).\label{integral3}
\end{eqnarray}
On each stratum $G_{n,2}^{[H]}$, stabilizer $H_{\mathcal{W}}$ is conjugate to a $H_0 \in [H]$ which does not depend on $\mathcal{W}$. Since the Haar measure $\mathrm{Vol}$ is invariant under conjugation, we have $\mathrm{Vol}(H_{\mathcal{W}}) = \mathrm{Vol}(H_0)$.

We denote $\mathrm{Vol}(H_0)$ by $v_{[H]}$ which is constant with respect to $\mathcal{W}$. Substituting into (\ref{integral2}) yields to
\begin{eqnarray}
	\int_{G_{n,2}} \log \mathrm{Vol}(H_{\mathcal{W}}) \mathrm{d}\mu(\mathcal{W})= \sum_{[H]\in \mathcal{S}}\log v_{[H]} \mu\left(G_{n,2}^{[H]}\right). \label{intfinal}
\end{eqnarray}
Substituting (\ref{intfinal}) into (\ref{integral2}) completes the proof.
\end{proof}

\begin{remark}
While the total symmetry $G$ governs the overall ambiguity, the constant $v_{[H]}$ accounts for the fact that some rotations in $G$ fix the chosen subspace $\mathcal{W}$ entirely (the stabilizer $H_{\mathcal{W}}$), thereby no contributing to the variability of the projections.
\end{remark}
We next describe how decrease in $\mathrm{Vol}(G)$ (more asymmetric $K_0$) causes increase in the mutual information $I(K_1;K_2|K_0)$. Using the chain rule, we have
\begin{eqnarray}
	&& \!\!\!\!\!\!\!\!\!\! h(K_1,\mathcal{W}_2|K_2,K_0) = h(\mathcal{W}_2|K_2,K_0)+h(K_1|\mathcal{W}_2,K_2,K_0)\nonumber\\
	&& \ \ \ \ \ \ \ \ \ \ \  = h(K_1|K_2,K_0) + h(\mathcal{W}_2|K_1,K_2,K_0). \label{49}
\end{eqnarray}
By (\ref{firstpart}) and (\ref{49}), it yields
\begin{eqnarray}
	&& \!\!\!\!\!\!\!\!\!\! h(K_1|K_2,K_0) = \mathbb{E}_{K_2|K_0}\left[\log N(K_0,K_2)\right] + \log \mathrm{Vol}(G_{n,2})\nonumber\\
	&& \ \ \ \ \ \ \  + h(K_1|\mathcal{W}_2,K_2,K_0) -h(\mathcal{W}_2|K_1,K_2,K_0). \label{50}
\end{eqnarray}
We wish to track the sensitivity of each term in $I(K_1;K_2|K_0) = h(K_1|K_0) - h(K_1|K_2,K_0)$ with respect to $\mathrm{Vol}(G)$ given (\ref{50}). Since $K_1$ is determined by the random direction $\mbox{\boldmath $U$}_{1} \sim \mathrm{Uniform}\left(\mathbb{S}^{n-1}\right)$ and is invariant under the finite symmetry group $G$ of $K_0$, the effective resulting space is the quotient manifold $\mathbb{S}^{n-1} / G$. The uniform distribution on the sphere pushes forward to a uniform distribution on quotient $\mathbb{S}^{n-1} / G$ with volume $\mathrm{Vol}(\mathbb{S}^{n-1}) / \mathrm{Vol}(G)$ (here $\mathrm{Vol}(\mathbb{S}^{n-1})$ denotes the surface of sphere), resulting into the entropy $h(K_1|K_0) = \log \mathrm{Vol}(\mathbb{S}^{n-1}) - \log \mathrm{Vol}(G)$. This yields a unit rate increase of $+1$ when $\mathrm{Vol}(G)$ decreases in $I(K_1;K_2|K_0)$. By Theorem \ref{ThmEBound}, the lower bound on $\mathbb{E}_{K_2|K_0}\left[\log N(K_0,K_2)\right]$ stems from the volume inequality $\mathrm{Vol}(G.\mathcal{W}) \leq \mathrm{Vol}(\mathcal{C}(K_0,K_2))$ which represents the measure of \textit{accidental} subspaces that happen to yield an identical projection without being related by a symmetry $g\in G$. For a generic convex body, this gap is negligible, implying that the ambiguity set $\mathcal{C}(K_0,K_2)$ is almost everywhere parameterized by $G/H_{\mathcal{W}}$. Thus, the bound in Theorem \ref{ThmEBound} achieves tightness, meaning a smaller $\mathrm{Vol}(G)$ decreases the subtracted $\mathbb{E}_{K_2|K_0}\left[\log N(K_0,K_2)\right]$, contributing a $+1$ rate increase to the mutual information. The other subtracted term $-\Delta h = -(h(K_1|\mathcal{W}_2,K_2,K_0) - h(\mathcal{W}_2|K_1,K_2,K_0))$ governs by $K_1$'s ambiguity orbit. The entropy $h(K_1|\mathcal{W}_2,K_2,K_0)$ is inversely proportional to the Haar measure of $K_1$'s stablizer $G_{K_1} = \{ g\in G: gK_1=K_1\} \subset G$, yielding $h(K_1|\mathcal{W}_2,K_2,K_0) \propto -\log \mathrm{Vol}(G_{K_1})$. Because $\mathrm{Vol}(G_{K_1}) < \mathrm{Vol}(G)$, we have $\alpha <1$ such that $\mathrm{Vol}(G_{K_1}) = \alpha \mathrm{Vol}(G)$. Consequently, $-\Delta h \propto \alpha \log \mathrm{Vol}(G)$. Therefore, the reduction in $\mathrm{Vol}(G)$ decreases $-\Delta h$, pulling the mutual information down by a rate $-\alpha$. Summing these variations yields a net sensitivity rate of $1+1-\alpha=2-\alpha >1$, ensuring that a reduction in $\mathrm{Vol}(G)$ increases the mutual information.

\section{Conclusion}

We studied dimensionality reduction via repeated random projections of a convex body, examining how much information is preserved at each step via mutual information. We show the projected bodies evolve in a Markovian manner and that greater shape asymmetry retains more information.

\section{Acknowledgement}
I would like to thank the anonymous reviewers for their time, effort, and constructive feedback, which greatly improved the clarity of this work.

\end{document}